\documentclass[aps,prb,showpacs,amsmath,twocolumn,superscriptaddress]{revtex4}

\usepackage{amsmath}
\usepackage{amssymb}
\usepackage{graphicx}
\usepackage{bbm}
\usepackage{bbold}

\newcommand{\vareps}{\varepsilon}

\newcommand{\vect}[1]{\mathbf{#1}}

\begin{document}

\title{ Monolayer $\textnormal{MoS}_2$: trigonal warping, ``$\Gamma$-valley'' and spin-orbit coupling effects}

\author{Andor Korm\'anyos}
\thanks{e-mail: andor.kormanyos@uni-konstanz.de}
\affiliation{Department of Physics, University of Konstanz, D-78464 Konstanz, Germany}

\author{Viktor Z\'olyomi}
\affiliation{Department of Physics, Lancaster University, Lancaster, LA1 4YB, United Kingdom}

\author{Neil D. Drummond}
\affiliation{Department of Physics, Lancaster University, Lancaster, LA1 4YB, United Kingdom}

\author{P\'eter Rakyta}
\affiliation{Budapest University of Technology and Economics, Department of Theoretical Physics 
and Condensed Matter Research Group of the Hungarian Academy of Sciences, 
Budafoki \'ut 8, H-1111 Budapest, Hungary}

\author{Guido Burkard}
\affiliation{Department of Physics, University of Konstanz, D-78464 Konstanz, Germany}

\author{ Vladimir I. Fal'ko}
\affiliation{Department of Physics, Lancaster University, Lancaster, LA1 4YB, United Kingdom}


\begin{abstract}
We use a combined \emph{ab-initio} calculations and $\vect{k}\cdot\vect{p}$ theory based approach to
derive a low-energy effective Hamiltonian for  monolayer $\textnormal{MoS}_2$ at the $K$ point of 
the Brillouin zone. It captures the 
features which are present in first-principles calculations but not explained by the 
theory of  Xiao \emph{et al}. [Phys Rev Lett \textbf{108}, 196802 (2012)], namely the 
trigonal warping of the valence and conduction bands, 
the electron-hole symmetry breaking, and the spin-splitting of the conduction band. 
We also consider other points in the Brillouin zone which might be important for transport properties. 
Our findings  lead to a more quantitative understanding of the properties of this material in the ballistic
limit.
\end{abstract}


\maketitle

\emph{Introduction}.
Transition metal dichalcogenides are emerging as promising new materials for applications in 
electronics and optoelectronics\cite{nnanotech-review}. 
In particular, monolayer molybdenum disulphide 
($\textnormal{MoS}_2$) has  recently received significant  attention 
experimentally\cite{kis,eda,cui,heinz,wang,wu,cao,buscema,sallen,xiaodong} as well as
theoretically\cite{yao,zhu,scuseria,lambrecht,niu,shen,jacobsen,ashwin,kosmider,yakobson,lee,hawrylak,wanxiang,sanvito,guinea,kim,song}.
It may become the material of choice for field-effect transistors with  high 
on-off ratio \cite{kis}. In addition, the strong spin-orbit coupling, the coupling between the spin and valley degrees 
of freedom\cite{yao,niu}, and their effect on the exciton photoluminescense have sparked strong 
interest\cite{eda,cui,heinz,wang,wu,cao,buscema,sallen,xiaodong}.  
In light of the growing interest to  this material, an accurate yet reasonably simple 
model that describes its band structure and electronic properties is highly desirable.

\begin{figure}[ht]
 \includegraphics[scale=0.5]{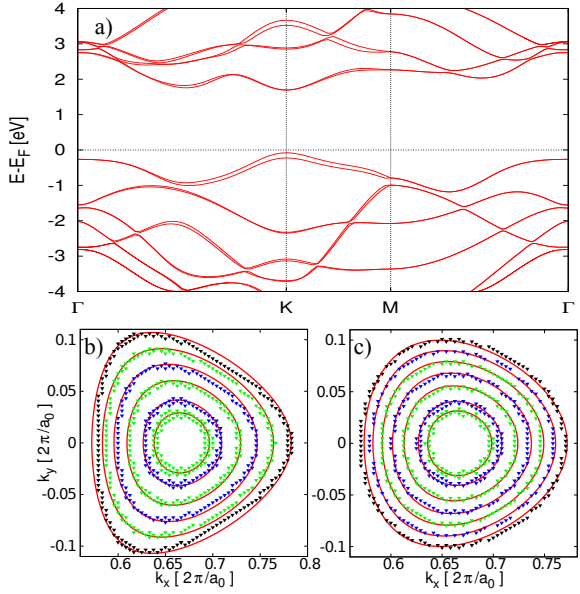}
 \caption{a) Spin-resolved band structure of $\textnormal{MoS}_2$ from DFT LSDA calculations. 
 b) Contour plot showing the isoenergy contours of the valence band (for zero SOC) from DFT calculations 
  at the $K$ point of the BZ (symbols) and  as  obtained from Eq.~(\ref{H_eff}) (solid lines). 
  c) the same as in b) for the conduction band. $a_0$ is the lattice constant.
     The energy difference between the
      two innermost contours is $0.02\,{\rm eV}$, between every other contours is $0.04\,{\rm eV}$.
   }
 \label{fig1}
\end{figure}

Following the important work in Refs.~[\onlinecite{yoffe}] and [\onlinecite{mattheis}],  
Xiao \emph{et al.}~\cite{yao} have recently
introduced a tight-binding model which assumes that at the $K$ point of the 
Brillouin zone (BZ) it is sufficient to take into account the $d_{z^2}$  (for the conduction band) and 
$d_{xy}$, $d_{x^2-y^2}$ (for the valence band) atomic orbitals of the Mo atoms. Neglecting 
the spin-orbit coupling (SOC),  Xiao \emph{et al.} 
found an effective Hamiltonian of the form
$
 \hat{H}_0=\hbar v_0 (\tau k_x\sigma_x+k_y\sigma_y)+\frac{\Delta}{2}\sigma_z
$
where $\sigma_{x,y,z}$ denote Pauli matrices, $\Delta$ is the energy gap, $v_0$ plays the role
of ``Fermi-velocity'' and $\tau=1\,(-1)$ for valley $K$ ($K'$).
$\hat{H}_0$ describes massive Dirac particles,  in other words, 
it is a monolayer graphene Hamiltonian with a staggered sublattice potential. 
While it 
seems to  explain many experimental observations at least qualitatively\cite{cao,cui,heinz}, 
certain limitations of this model  can already be appreciated by looking at Figs.~\ref{fig1}(b) and (c). 
The dispersion predicted by $\hat{H}_0$ is isotropic and possesses 
electron-hole symmetry regarding the  valence and conduction bands.
As one can see in Figs.~\ref{fig1}(b) and (c), which show the results of first-principles calculations,
except in the immediate vicinity of the K point the dispersion is 
not isotropic: a trigonal warping (TW) of the isoenergy contours can  clearly be seen.
In comparison  to monolayer graphene we note that its low-energy dispersion is isotropic on the energy scale 
of $1\, {\rm eV} $, whereas in $\textnormal{MoS}_2$ the TW 
is already observable at $\approx 0.08\, {\rm eV}$ below the valence-band edge. 
Furthermore, \emph{ab initio} calculations predict different effective 
masses for electrons and holes\cite{lambrecht,ashwin,yakobson}, which obviously breaks the 
electron-hole symmetry. 
We also note that both our density functional  (DFT) calculations  
and the  computations of Refs.~[\onlinecite{hawrylak,lambrecht, kosmider}]
indicate that there is a relatively small (compared to the corresponding splitting in the valence band) but finite 
spin-splitting of about 3--4 meV  in the conduction band at the $K$ point, 
which cannot be explained in the theoretical framework of Ref.~[\onlinecite{yao}].
Finally, as one can observe  in Fig.~\ref{fig1}(a) (see also  the  calculations of  
Refs.~[\onlinecite{scuseria,lambrecht,yakobson}]) the valence-band maximum (VBM) at the $\Gamma$ point is actually 
 very close in energy to the valence-band  edge 
 at the $K$ point. While the exact value of the band maximum seems to depend on the particular computational method
 that is used (c.f. Fig.~1(a) in Ref.~[\onlinecite{lambrecht}] and Fig.~3 in Ref.~[\onlinecite{yakobson}]), 
 it is clear that at finite temperatures  in hole-doped samples  states at both  $K$ and  $\Gamma$ 
 points will participate in  transport (for other points of interest in the BZ see Appendix \ref{sec:qpoint}). 
These observations call for a more exact model for the band structure of   $\textnormal{MoS}_2$.

Using our \emph{ab initio} computations and the $\mathbf{k}\cdot\mathbf{p}$ theory\cite{dresselhaus-book} we show 
that the observed TW of the bands can be  described by a four-band  generalized bilayer-graphene-type Hamiltonian
and the TW 
is understood as a consequence of the coupling of the valence band (VB) and conduction band (CB)
to other (remote) bands. From the four-band model we derive an effective two-band model and obtain the parameters that enter
the model from fitting to our DFT computations.  SOC also plays an important role in the low-energy physics of 
$\textnormal{MoS}_2$, 
but the SOC Hamiltonian  cannot be obtained from results on bilayer graphene\cite{ABC_SO,fabian_bilayer_SO}. 
We find that a  consistent description requires a seven-band model (or fourteen-band model, including the spin) at the 
$K$ point and a six-band model at the $\Gamma$ point.  We derive an effective low-energy Hamiltonian which 
takes into account the effects of the SOC, including the spin-splitting of the CB, which, to our knowledge
has not been considered before. We also
discuss how the effective mass and various other parameters of the model depend on the SOC.

\emph{Theory and results}.
We start  with the derivation of the spinless  effective Hamiltonian, i.e. we neglect the SOC. 
This is already sufficient to  explain the TW 
of the isoenergy contours.  
We  make repeated use  of various symmetries of the crystal structure, the two most important of which   
for our  purposes are the rotational symmetry  by $2\pi/3$ around an axis perpendicular to the plane of  $\textnormal{MoS}_2$ 
(we denote it by $C_3$) and the horizontal mirror plane $\sigma_h$.  
(For the full point group symmetry see Appendix \ref{sec:character-tables}.)
The derivation of  the effective Hamiltonian relies on our DFT calculations, which, in addition to the band structure, 
provide us with the projection of the DFT wave functions  onto atomic orbitals at high symmetry points of the BZ. 
This helps us to identify  the symmetries of the bands, which is necessary to obtain the general 
form of the effective Hamiltonian. As an example we consider the (topmost) valence and the 
(lowest) conduction bands at the $K$ point of the BZ. 
Similarly to Refs.~[\onlinecite{scuseria,yao,yakobson,hawrylak}], we  find that here the VB 
is predominantly composed of the $d_{x^2-y^2}$ and $d_{xy}$ atomic orbitals centered
on the Mo atoms, which are symmetric with respect to $\sigma_h$. 
Since the VB is  non-degenerate at the 
 $\Gamma$ point, compatibility relations require that at the $K$ point it transforms as the $A'$  representation of 
 the group $C_{3h}$, which is  the small group of the wave vector at $K$. 
 We denote the wavefunction of the VB by $|\Psi^{v}_{A'}\rangle$, 
 and  hereafter we use the notation $|\Psi^{b}_{\mu}\rangle$ for the wavefunctions of various bands,
where $b$ denotes the band and $\mu$ the pertinent irreducible representation (irrep).
The CB is predominantly composed of $d_{z^2}$ orbitals of the Mo atoms\cite{scuseria,yao,yakobson,hawrylak}, 
which means that the CB wavefunction  
$|\Psi^{c}_{E_1'}\rangle$ is also symmetric with respect to $\sigma_h$  and 
transforms as the $E_1'$ irrep of $C_{3h}$.
Similar considerations allow us to obtain the symmetries of all bands
at the $K$ point, even when different orbitals from different atoms are admixed  
(see Appendix \ref{sec:character-tables}). 
In a minimal model that captures TW, in addition to the VB and the CB,  
there are two other important bands, both of which are
even with respect to $\sigma_h$: 
the second one above the CB, which we denote by CB+2 and whose wavefunction is $|\Psi^{c+2}_{E_2'}\rangle$ 
and the third one below the valence 
band (VB-3) with wavefunction $|\Psi^{v-3}_{E_2'}\rangle$. 
The  other bands between the VB-3 and  CB+2 are antisymmetric with respect to 
the mirror plane of $\textnormal{MoS}_2$ and therefore they do not couple to the VB and the CB. 
In $\mathbf{k}\cdot\mathbf{p}$ theory the Hamiltonian 
$
\mathcal{H}_{\mathbf{k}\cdot\mathbf{p}}
=\frac{\hbar}{m_e}\mathbf{k}\cdot\hat{\mathbf{p}}
$ 
is considered a perturbation ($m_e$ is the bare electron mass) and one uses first-order perturbation theory  
in the basis of 
$\{|\Psi_{A'}^{v}\rangle,\,|\Psi_{E_1'}^{c}\rangle, \, |\Psi_{E_2'}^{v-3}\rangle,\, |\Psi_{E_2'}^{c+2}\rangle \}$.  
Using the notation $\mathbf{k}=\mathbf{q}+\mathbf{K}$, the perturbation  
can be rewritten as 
$
\mathcal{H}_{\mathbf{k}\cdot\mathbf{p}}=\frac{1}{2}\frac{\hbar}{m_e} (q_{+} \hat{p}_{-} + q_{-} \hat{p}_{+})
=\mathcal{H}_{\mathbf{k}\cdot\mathbf{p}}^{-} + \mathcal{H}_{\mathbf{k}\cdot\mathbf{p}}^{+},
$ 
where the operators  $\hat{p}_{\pm}$ are defined as 
 $\hat{p}_{\pm}=\hat{p}_x\pm i \hat{p}_y$ and similarly $q_{\pm}=q_x\pm i q _y$. 
 The matrix elements of $\mathcal{H}_{\vect{k}\cdot\vect{p}}$
 are constrained by the symmetries of the system. For example, 
 considering the rotation  $C_3$,   the relation  
 $
 \langle \Psi_{A'}^{v}| \hat{p}_{+} | \Psi_{E_2^{'}}^{c+2}\rangle=
 \langle \Psi_{A'}^{v}| C_3^{\dagger} C_3 \,\hat{p}_{+}\, C_3^{\dagger} C_3 |  \Psi_{E_2'}^{c+2}\rangle
 $ 
 should hold. Since 
 $\langle \Psi_{A'}^{v}| C_3^{\dagger}= \langle \Psi_{A'}^{v}|$, 
 $C_3 \hat{p}_{\pm} C_3^{\dagger}=e^{\mp i 2 \pi/3}\hat{p}_{\pm}$ and 
 $ C_3 |\Psi_{E_2'}^{c+2}\rangle = e^{ -i 2 \pi/3}|\Psi_{E_2'}^{c+2}\rangle $ one obtains that
$
\langle \Psi_{A'}^{v}| \mathcal{H}_{\mathbf{k}\cdot\mathbf{p}}^{+} | \Psi_{E_2'}^{c+2}\rangle= 
e^{ -i 4 \pi/3} \langle \Psi_{A'}^{v}| \mathcal{H}_{\mathbf{k}\cdot\mathbf{p}}^{+} | \Psi_{E_2'}^{c+2}\rangle,
$ 
which means that this matrix element must vanish. By contrast,  
$\langle \Psi_{A'}^{v}| \hat{p}_{-} | \Psi_{E_2'}^{c+2}\rangle=\gamma_4$
is finite
and one can prove that it is a real number. Similar considerations can be used to calculate all matrix elements. 
Finally, in the basis of 
$\{|\Psi_{A'}^{v}\rangle,\,|\Psi_{E_1'}^{c}\rangle, \, |\Psi_{E_2'}^{v-3}\rangle,\, |\Psi_{E_2'}^{c+2}\rangle \}$ 
the $\mathbf{k}\cdot{\mathbf{p}}$ Hamiltonian  at the $K$ point is  given by
\begin{equation}
{H}_{\mathbf{k}\mathbf{p}}=
\left(
\begin{array}{cccc}
 \vareps_{v} & \gamma_3 q_{-} & \gamma_2 q_{+} & \gamma_4 q_{+} \\
 \gamma_3 q_{+} & \vareps_c & \gamma_5 q_{-} & \gamma_6 q_{-}\\
 \gamma_2 q_{-} & \gamma_5  q_{+} & \vareps_{v-3} & 0 \\
 \gamma_4  q_{-} & \gamma_6  q_{+} & 0 & \vareps_{c+2}\\ 
\end{array}
\right),
\label{HkpK}
\end{equation}
where $\gamma_i$ are the matrix elements of $\mathcal{H}_{\vect{k}\cdot\vect{p}}$ in the above mentioned basis and 
$ \vareps_{v},  \vareps_c, \vareps_{v-3}, \vareps_{c+2}$ are band-edge energies. 
The matrix element between $|\Psi_{E_2'}^{v-3}\rangle$ and  $|\Psi_{E_2'}^{c+2}\rangle$ vanishes due to
symmetry.
We note in passing that the Hamiltonian (\ref{HkpK}) can be considered a generalized bilayer 
graphene (BLG) Hamiltonian\cite{mccann}.
This can be seen by rotating the well known bilayer graphene Hamiltonian into a basis where the basis functions
transform according to the irreps of the small group of BLG (see Appendix \ref{sec:bilayer}). 
 To obtain the  $\mathbf{k}\cdot{\mathbf{p}}$ Hamiltonian  at the $K'$ point it proves to be useful to re-define 
$q_{\pm}$ as $q_{\pm}=q_x\pm i\,\tau\,q_y$, where $\tau=1\, (-1)$ for the $K$ ($K'$) point.
As in the case of BLG, it is  convenient to derive a low-energy effective Hamiltonian from (\ref{HkpK}), which does 
not contain the contribution of the bands far from the Fermi energy.  
Using the L\"owdin partitioning\cite{winkler-book} we find 
\begin{subequations}
 \begin{eqnarray}
{H}_{\rm eff} &=& {H}_0 + {H}_{as} + {H}_{3w} + H_{cub},\\
{H}_0 + {H}_{as} &=& 
\left(
\begin{array}{cc}
 \vareps_v & \tau\,\gamma_3 q_{-} \\
 \tau\,\gamma_3 q_{+} & \vareps_c\\
\end{array}
\right) + 
\left(
\begin{array}{cc}
 \alpha q^2 & 0 \\
 0  & \beta q^2 \\
\end{array}
\right),\\ 
{H}_{3w} &=& \kappa
\left(
\begin{array}{cc}
 0 &   (q_{+})^2 \\
  (q_{-})^2 & 0\\
\end{array}
\right),\\
H_{cub} &= & -\tau\,\frac{\eta}{2} q^2 
\left(
\begin{array}{cc} 
 0 &   q_{-} \\
  q_{+} & 0\\
\end{array}
\right),
\end{eqnarray}
\label{H_eff}
\end{subequations}\noindent
where the parameters $\alpha$, $\beta$, $\kappa$ and $\eta$ can be expressed in terms of 
$\gamma_i$ as defined in Eq.~(\ref{HkpK}) and 
the band edge energies.  The parameters $\alpha$ and $\beta$ describe the breaking of the 
electron-hole symmetry that is 
apparent comparing Figs.~\ref{fig1}(b) and (c), whereas $\kappa$ is responsible for the TW 
of the energy contours. The cubic term, $H_{cub}$ is important to achieve a quantitative fit to the VB away from the $K$ point.
We note that symmetries allow for another, diagonal Hamiltonian which is $\sim q^3$, but its effect has been 
found to be negligible.   

We used the VASP code\cite{vasp}  and the local density approximation (LDA) of 
DFT to calculate\cite{viktor} the band structure and isoenergy contours shown in Fig.~\ref{fig1}.  
To compare our $\mathbf{k}\cdot\mathbf{p}$ theory with the DFT calculations one has to determine
the matrix elements $\gamma_i$. These matrix elements, in principle, 
can also be calculated from DFT\cite{shou-cheng}. 
We found however, that the isoenergy lines calculated with parameters $\gamma_i^{\rm DFT}$ 
obtained from the numerical evaluation of $\langle\Psi_{\mu}^{b}|\hat{p}_{\pm}|\Psi_{\mu'}^{b'}\rangle$ 
using Kohn-Sham orbitals\cite{neil-derivative} give 
a satisfactory agreement with DFT band structure only in the close vicinity of the $K$ point. 
Therefore we  used  these $\gamma_i^{\rm DFT}$  values as a starting point for a fitting procedure 
whereby we fitted the eigenvalues  of the Hamiltonian (\ref{H_eff}) 
[given in terms of $\gamma_3,\alpha,\beta, \kappa$, $\eta$ and the band edge energies 
$\vareps_v$, and $\vareps_c$] along the $\Gamma K M$ line 
to the CB and VB obtained from the DFT calculations. 
The fit involved a range of $\approx 0.1\times \frac{2\pi}{a_0}$ in the $\Gamma$ and $M$ directions.
From the fitting we  found  the parameters 
$\gamma_3=3.82\,{\rm eV \AA{}}, \alpha=1.72\,{\rm eV \AA{}^2}, \beta=-0.13\, {\rm eV \AA{}^2}$,  
$\kappa=-1.02\,{\rm eV \AA{}^2}$ and $\eta = 8.52\, {\rm eV \AA{}^3}$.
The isoenergy contours calculated using these parameters 
[the solid lines in Figs.~\ref{fig1}(b) and (c)] capture  well the TW 
of the band dispersion (c.f the results of the DFT computations given by symbols), which is more pronounced 
in the VB than  in the CB. The agreement between the DFT results and the predictions based on (\ref{H_eff}) is
very good up to energies  $0.16\,{\rm eV}$ below the VB maximum and  above the CB minimum;
for other energies the agreement is  qualitative.
The effective masses (along the $\Gamma K M $ line) that can be inferred from these parameters 
are $m_{\rm eff}^{v}\approx -0.62 m_e$ for the VB
and $m_{\rm eff}^{c}\approx 0.48 m_e$ for the CB, which are in good agreement with the results of 
other DFT LDA calculations\cite{jacobsen,lee}. (For a similar set of parameters based on band structure calculations 
using the HSE06 hybrid functional, see Appendix \ref{hse-bandparam}).
Interestingly, we have checked by numerical calculations that although the TW is quite strong in the 
VB, its effect on the Landau levels is actually very small,  and they can be calculated by neglecting $H_{3w}$
in Eq.~(\ref{H_eff}). Nevertheless, $H_{3w}$ and $H_{cub}$ should affect the Berry-curvature and hence 
various Hall-conductivities\cite{yao,wanxiang}. 

An important feature of the band structure of $\textnormal{MoS}_2$ which has received little attention so far 
is that the top of the valence band at the $\Gamma$ point is very close in energy to the VBM
at the $K$ point\cite{scuseria,lambrecht,yakobson}; see also Fig.\ref{fig1}(a). 
This means that for the VB transport properties 
the states close to the $\Gamma$ point can  also be important. 
The Hamiltonian of this ``$\Gamma$-valley'' can also be derived using the $\mathbf{k}\cdot\mathbf{p}$ theory, 
along similar lines to the case of the $K$ point. Note, however, that 
the group of the wave vector at the $\Gamma$ point is $D_{3h}$. 
Our DFT calculations show that here the VB 
is mainly composed of the $d_{z^2}$ and $s$ orbitals of the Mo atoms and the $p_z$ orbitals of the S atoms, 
which means that it belongs to the  $A_1^{'}$ irrep of $D_{3h}$. 
The VB is coupled to the VB-3 and  CB+1 bands
which are doubly degenerate at the $\Gamma$ point.
There is no coupling between the VB and the CB at the $\Gamma$ point: 
due to band crossings along the $\Gamma-K$ line the CB becomes antisymmetric 
with respect to $\sigma_h$. 
Upon performing the L\"owdin partitioning we find that 
the dispersion is isotropic and can be well described by $\mathcal{H}_{\Gamma}=\frac{\hbar^2 k^2}{2 m_{\rm eff}^{\Gamma}}$, 
where the effective mass $m_{\rm eff}^{\Gamma}=-3.65 m_e$ is found by fitting the band structure, which is in
good agreement with Ref.~[\onlinecite{lee}]. 
The importance and  role of the $\Gamma$ point in the transport properties of the VB would require further 
experimental work.
We expect  that in disordered samples due to their large effective mass and hence low mobility 
the contribution of these states to the transport is small, but they 
may be important in ballistic samples and in quantum-Hall measurements.

The description of the system becomes more complicated if one takes into account the  SOC
as well. In the atomic approximation the SOC is given by the Hamiltonian
\begin{equation}
 \mathcal{H}_{\rm so}^{\rm at}=\frac{\hbar}{4 m_e^2 c^2} \frac{1}{r} \frac{d V(r)}{d r} \,\mathbf{L} \cdotp \mathbf{S}.
 \label{atomic-SOC}
\end{equation}
Here $V(r)$ is the spherically symmetric atomic potential, $\mathbf{L}$ is the  angular momentum operator and 
$\mathbf{S}=(S_x,S_y, S_z)$ is a vector of spin Pauli matrices $S_x,\,S_y$ (with eigenvalues $\pm 1$).
Note that $\mathbf{L} \cdotp \mathbf{S}=L_z S_z + L_{+} S_{-} + L_{-} S_{+}$, where $L_{\pm}= L_{x} \pm i L_{y}$ and 
$S_{\pm}=\frac{1}{2}(S_{x}\pm i S_{y})$. 
Let us introduce the spinful symmetry basis functions by 
$| \Psi_{\mu}^{b}, s\rangle =  | \Psi_{\mu}^{b}\rangle \otimes | s \rangle$, 
where $s=\{ \uparrow,\downarrow \}$ denotes the spin degree of freedom, and 
consider first the $K$ point of the BZ\@. 
Since $L_{\pm}$ transforms as the $E''$ irrep of $C_{3h}$,  there can be 
non-vanishing matrix elements of $\mathcal{H}_{\rm so}^{\rm at}$ between states that are even/odd with respect to 
$\sigma_h$. Therefore we considered a seven-band model (without spin) 
which contains every band between VB-3 and CB+2, i.e., we consider the basis 
$\{ |\Psi_{E_2^{'}}^{v-3},s\rangle, |\Psi_{E_1^{''}}^{v-2},s\rangle, |\Psi_{E_2^{''}}^{v-1},s\rangle, 
|\Psi_{A'}^{v},s\rangle, |\Psi_{E_1^{'}}^{c},s\rangle, |\Psi_{A^{''}}^{c+1},s\rangle, \\
|\Psi_{E_1^{'}}^{c+2},s\rangle
\}$. 
The  symmetries $\sigma_h$ and $C_3$ of the system here also help us to find the non-zero matrix elements 
of  $\mathcal{H}_{\rm so}^{\rm at}$. For example,  one can make use of the fact that 
$
C_3\,L_{\pm}\,C_{3}^{\dagger}=e^{\mp i 2\pi/{3}} L_{\pm}
$
and therefore show that $\langle s,\Psi_{A'}^{v}| L_{-}S_{+} |\Psi_{E_2^{''}}^{v-1},s \rangle
=\Delta_{(v,v-1)} S_{+}$ 
where $\Delta_{(v,v-1)}$ is a constant, whereas  
$\langle s,\Psi_{A'}^{v}|L_{+} S_{-}|\Psi_{E_2^{''}}^{v-1},s\rangle= 
\langle s,\Psi_{A'}^{v}|L_{z} S_{z}|\Psi_{E_2^{''}}^{v-1},s\rangle=0$.
By calculating the matrix ${H}_{so}^{at}$ in the above mentioned basis and diagonalizing the 
Hamiltonian $H_{d}+H_{so}^{at}$ where $H_{d}$ is a diagonal matrix containing the band-edge energies, 
one  obtains the eigenstates $| \Psi_{\mu,\mu'}^{b}, s \rangle$, 
which in general turn out to be linear combinations of 
a symmetric $|\Psi_{\mu}^{b},s\rangle$ and an antisymmetric $|\Psi_{\mu'}^{b'},s\rangle$ 
wavefunction with different weights. 
In our notation  the new eigenstates 
$| \Psi_{\mu,\mu'}^{b}, s \rangle$ inherit the band index $b$ and 
spin index $s$ from the state whose weight is larger in the linear combination that makes 
up $| \Psi_{\mu,\mu'}^{b}, s \rangle$. 
This assignment of the band index and spin quantum number is possible because the typical energy scale of the 
SOC (the upper limit of which is  the splitting of the valence band $\approx 145 {\rm meV}$, see below) 
is significantly smaller than the typical band separation, i.e., the bands are not strongly hybridized by the SOC. 
The diagonalization of the Hamiltonian can be done analytically in the approximation that 
couplings of  up to next-nearest-neighbour bands are kept and more remote couplings,
e.g., between $|\Psi_{E_2^{'}}^{v-3},s\rangle$ and   $|\Psi_{A^{''}}^{c+1},s\rangle$
are neglected. All eigenstates are non-degenerate, as expected, since 
the double group of $C_{3h}$ has only one-dimensional representations. 
With the new eigenstates  $| \Psi_{\mu,\mu'}^{b}, s \rangle$  one can repeat the 
$\mathbf{k}\cdot\mathbf{p}$ calculation, and since $|\Psi_{\mu,\mu'}^{b}, s \rangle$ is an 
 admixture of symmetric and antisymmetric states, there will be more non-zero matrix elements of the 
 $\mathcal{H}_{\mathbf{k}\cdot\mathbf{p}}$ Hamiltonian than there  were in the case of zero spin-orbit 
 coupling; see Eq.(\ref{HkpK}).  Finally, using the L\"owdin partitioning 
 we can derive an effective low-energy Hamiltonian for
 the spinful valence and conduction bands. 
Since the calculations are quite lengthy, 
we only give 
the most important results here and concentrate on the zero-magnetic field case. 
The Landau-level problem in the presence of SOC and the related question of 
the  effective $g$-factor of monolayer $\textnormal{MoS}_2$ will be discussed elsewhere.

We will work in the basis of 
$\{|\Psi^{v}_{A',E_2^{''}},\uparrow\rangle, |\Psi^{v}_{A',E_1^{''}},\downarrow\rangle, \\
|\Psi^{c}_{E_2^{'},E_{1}^{''}},\uparrow\rangle, |\Psi^{c}_{E_2^{'}, A^{''}},\downarrow\rangle\}$
and start with the diagonal and 
${\mathbf{q}}$ independent part of the  SOC Hamiltonian, i.e., we consider the SOC dependence of the
band edge energies.
According to our $\vect{k} \cdot \vect{p}$  calculations, the spin-splitting in the  VB and the CB can be
described by the Hamiltonians 
\begin{subequations}
\begin{eqnarray}
 H^{so}_{vb} &=& -\tau \Delta_{1}^{v} S_z + \frac{\Delta^v_2}{2}(\mathbb{1}+\tau S_z),
 \label{so-vb}\\
H^{so}_{cb}  &=&  \frac{|\Delta^{c}|^2}{2} \left[ \frac{\mathbb{1}- \tau S_z}{\vareps_c-\vareps_{v-1}}
 -\frac{\mathbb{1} + \tau S_z}{\vareps_{c+1}-\vareps_c}\right] \label{so-cb}.
 \end{eqnarray}
\end{subequations}
The term  $ -\tau \Delta_{1}^{v} S_z$  was first obtained in Ref.~\onlinecite{yao}, 
 whereas the second term of $ H^{so}_{vb}$, which is expected to be much smaller, comes from the coupling of spin-up 
 (spin-down) band  of VB to VB-1 at the $K$ ($K'$) point. 
 (The coupling of the spin-down (spin-up) band of the VB to other bands is much weaker.)
Our DFT calculations give a spin-orbit gap of $2 \Delta^{v}_{1} - \Delta_{2}^{v} \approx 146 {\rm meV}$ in the VB.
The  spin splitting of the CB, given by Eq.~(\ref{so-cb}),  although noted in 
Refs.~\onlinecite{hawrylak,lambrecht,kosmider}, has not yet been discussed in the literature.
It originates from the SOC of the CB to the VB-1 and CB+1 bands  and is  a consequence of 
 the hitherto neglected off-diagonal SOC terms, related to the $\sim L_{-}S_{+}+L_{+}S_{-}$ part of
 $\mathcal{H}_{\rm so}^{\rm at}$. Our results therefore show that the spin-valley coupling is present not only
 in the VB\cite{yao} but also in the CB. Our DFT computations give a spin-splitting of 
 $ 
 |\Delta^{c}|^2
 \left[\frac{1}{\vareps_c-\vareps_{v-1}}
 +\frac{1}{\vareps_{c+1}-\vareps_c}\right]
 \approx 3 {\rm meV}
$. 
Although this is a small effect compared to the spin-splitting in the VB,  spin-splittings  of similar magnitude have recently been
 measured in, e.g., carbon nanotube quantum dots\cite{rasmussen}.

Regarding the effect of SOC on the $\mathbf{q}$-dependent terms in Eq.~(\ref{H_eff}), we find that 
bands with different spin indices,
e.g., $\{|\Psi^{v}_{A',E_2^{''}},\uparrow\rangle$ and  $|\Psi^{v}_{A',E_1^{''}},\downarrow\rangle$ or 
 $ |\Psi^{c}_{E_2^{'}, A^{''}},\downarrow\rangle\}$ do not couple to each other. After folding down the 
 full seven-band $\vect{k} \cdot \vect{p}$ Hamiltonian,  in the basis of  
$\{|\Psi^{v}_{A',E_2^{''}},\uparrow\rangle, \\ |\Psi^{c}_{E_2^{'},E_{1}^{''}},\uparrow\rangle\}$ and 
$\{|\Psi^{v}_{A',E_1^{''}},\downarrow\rangle,|\Psi^{c}_{E_2^{'}, A^{''}},\downarrow\rangle\}$ 
the effective Hamiltonian is still of the form of Eq.~(\ref{H_eff}), but in general with different parameters 
$\gamma_3^{\uparrow (\downarrow)}$, $\alpha^{\uparrow (\downarrow)}$, $\beta^{\uparrow (\downarrow)}$,
$\kappa^{\uparrow (\downarrow)}$ and $\eta^{\uparrow (\downarrow)}$ for the spin-up (spin-down) bands.
 By fitting our SOC-resolved DFT calculations
 we find that  $\gamma_3$ and $\beta$, hence  $m_{\rm eff}^c$ are basically not affected by the SOC. 
 The effective masses in the VB are slightly renormalized by the SOC,  leading to 
 $m_{\rm eff}^{v, \uparrow}\approx 0.65 m_e$ and $m_{\rm eff}^{v, \downarrow}\approx 0.58 m_e$, i.e.,  
a difference of roughly $5\%-6\%$ with respect to the zero SOC case.

Finally, we briefly discuss the effect of  SOC on the states at the $\Gamma$ point of the BZ. 
In contrast to the $K$ point, here the valence band remains degenerate even if we take into account SOC (see Fig.\ref{fig1}). 
This can be understood from general group theoretical arguments: the pertaining double group has two-dimensional,
hence degenerate representations. The dispersion for each spin can be described by a parabolic dependence on $\mathbf{k}$
and we find that the effective mass is basically unchanged with respect to the zero SOC case.

\emph{Conclusions}. We have derived a low-energy effective Hamiltonian for monolayer 
$\textnormal{MoS}_2$ at the $K$ point of the BZ, which takes into account effects that are present in 
first-principles calculations but have not hitherto been discussed.  
Our theory is valid at low densities and for perfectly flat monolayer $\textnormal{MoS}_2$ crystals.
The TW and spin-splitting 
of the bands should be readily observable by spin and angle resolved photoemission spectroscopy.
We have also considered the states at the $\Gamma$ point of the BZ, which can be important for transport 
properties of hole-doped samples as well as for various scattering and relaxation processes\cite{song,kim},
because scattering from the $K$ to the $\Gamma$ point does not require a simultaneous valley- and spin-flip.

\emph{Acknowledgements}.
A. K. and G. B. acknowledge funding from  DFG under programs SFB767, SPP1285, and FOR912.
V. Z.  acknowledges  support from the Marie Curie project CARBOTRON.

\emph{Note added}. During the preparation of this manuscript two  related preprints  have  appeared\cite{asgari,ochoa}, 
where some of the results that we present here have also been obtained.

\appendix

\section{Character tables and basis functions}
\label{sec:character-tables}

\begin{figure}[hbt]
\includegraphics[scale=0.25]{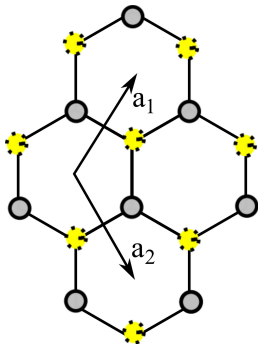}
\caption{Top view of the $\textnormal{MoS}_2$ lattice. Mo atoms are indicated by grey (solid line) circles,
S atoms by yellow (dotted line) circles.  The lattice vectors 
$\mathbf{a}_1=\frac{a_0}{2}(1,\sqrt{3})$ and  $\mathbf{a}_2=\frac{a_0}{2}(1,-\sqrt{3})$ are also shown
($a_0=3.129$\AA \,\,is the lattice constant).
\label{fig:mos2}}
\end{figure}

In Fig.~\ref{fig:mos2} we show a top view of the monolayer $\textnormal{MoS}_2$ lattice.
The pertinent point groups to  understand the band structure of monolayer $\textnormal{MoS}_2$ are  
$D_{3h}$ and $C_{3h}$. The former is the group of the wave vector  at the $\Gamma$ point, the latter at the $K$ point of 
the Brillouin zone (BZ). The symmetry operations that generate these groups are three-fold rotation $C_3$ 
around an axis perpendicular to the plane of $\textnormal{MoS}_2$, a horizontal mirror plane $\sigma_h$ perpendicular to
the three-fold axis and in the case of  $D_{3h}$, three two-fold rotation axis $C_2'$ that lie
in  the horizontal mirror plane. 

By projecting the plane-wave basis used in our DFT computations  onto atomic orbitals one can obtain the 
decomposition of each band in terms of atomic orbitals $\Phi^{\eta}_{\nu}$, where 
$\eta=\{\textnormal{Mo},\textnormal{S1}, \textnormal{S2} \}$ 
denotes whether the given orbital is centered on molybdenum (Mo) or on one of the sulphur 
(S1, S2) atoms in the unit cell, and the lower index 
$\nu=\{s,p_x,p_y,p_z,d_{z^2},d_{xy},d_{xz},d_{xy},d_{x^2-y^2}\}$ indicates the type of orbital.
To take into account the three-fold rotational symmetry of the system, 
one should use  linear combinations of 
 these orbitals to form the rotating orbitals $Y_{l}^{m}$,   which are proportional to spherical harmonics. 
We then consider  the transformation properties of the Bloch wave functions formed with the rotating orbitals: 
\begin{equation*}
 |\Psi_{l,m}^{\eta}(\vect{k},\vect{r})\rangle=\frac{1}{\sqrt{N}}
 \sum_n e^{i \vect{k}\cdot(\vect{R}_n+\vect{t}_{\eta})} \, Y_{l}^{m}(\vect{r}-[\vect{R}_n+\vect{t_{\eta}}]).
\end{equation*}
Here the summation runs over all lattice vectors $\vect{R}_n$ and $\vect{t}_{\rm Mo}=\frac{a_0}{2}(1,-\frac{1}{\sqrt{3}})^{T}$, 
$\vect{t}_{\rm S1}=\vect{t}_{\rm S2}=\frac{a_0}{2}(1,\frac{1}{\sqrt{3}})^{T}$ give the position of the Mo and S atoms in the  
(two-dimensional) unit cell with $a_0=3.129\AA$ being the lattice constant (see also Fig.\ref{fig:mos2})
and $\mathbf{k}$ is measured from the $\Gamma$ point of the BZ.  
We then identify the irreducible representations (irreps) according to which $|\Psi_{l,m}^{\eta}(\vect{k},\vect{r})\rangle$ 
transform at the high symmetry points $\Gamma$ and $K$ of the BZ. 
Since hybridization between different orbitals will preserve the symmetry properties, 
the analysis of the bands in terms 
of atomic orbitals, together with band compatibility relations, gives us the irreps that 
can be assigned to each band. 

As an example we consider the  valence  band (VB) at the $K$ point. 
Here the VB  is  predominantly composed of the $d_{x^2-y^2}$ and $d_{xy}$ atomic orbitals centered
on the Mo atoms, which are symmetric with respect to $\sigma_h$.
 The two Bloch functions that can be formed from these orbitals and which reflects the three-fold rotational 
 symmetry are 
 $
 |\Psi^{\rm Mo}_{2,\pm 2}(\mathbf{k})\rangle=\frac{1}{\sqrt{2 N}} \sum_{n} e^{i \mathbf{k}(\mathbf{R}_n+\mathbf{t}_{\rm Mo})} 
 Y_{2}^{\pm 2}(\mathbf{r}-(\mathbf{R}_n+\mathbf{t}_{\rm Mo}))
 $
 where 
 $Y_{2}^{\pm 2}(\mathbf{r})\sim \left(d_{x^2-y^2}(\mathbf{r}) \pm i d_{xy}(\mathbf{r}) \right)/\sqrt{2}$. 
 The small group of the wave vector at the $K$ point is $C_{3h}$.  One can show that
  $|\Psi^{\rm Mo}_{2,2}(\mathbf{K})\rangle$  transforms as the $E_2'$ 
 irrep of this group whereas  $|\Psi^{\rm Mo}_{2,-2}(\mathbf{K})\rangle$  transforms as the $A'$ irrep.  
Since the VB is  non-degenerate at the 
 $\Gamma$ point, compatibility relations require that at the $K$ point it transforms as the $A'$  irrep.

Table~\ref{tbl:D3h-group} shows the characters and irreps for $D_{3h}$, pertinent to the $\Gamma$ point, 
while Table~\ref{tbs:bandsymGamma} shows symmetry properties of the rotating orbitals based  Bloch wave functions  
and that which band they contribute to. The conduction band is denoted by CB, the valence by VB,  
the first band above the CB  by CB+1, the first band below the VB by VB-1, and so on.

\begin{table}[htb]
\begin{tabular}{|c|cccccc|}\hline
 $ \overline{6}m2\, (D_{3h})$ &  $E$  & $\sigma_h$ & $2 C_3$  &  $2 S_3$  &  $3 C_2'$ & $3\sigma_v$  \\
 \hline
 $A_1'$  & $1$  & $1$  & $1$  & $1$  & $1$  & $1$\\ 
 $A_2'$  & $1$  & $1$  & $1$  & $1$  & $-1$ & $-1$\\
 $A_1''$ & $1$  & $-1$ & $1$  & $-1$ & $1$  & $1$\\
 $A_2''$ & $1$  & $-1$ & $1$  & $-1$ & $-1$ & $-1$\\
 $E'$    & $2$  & $2$  & $-1$ & $-1$ & $0$  & $0$\\
 $E''$   & $2$  & $-2$ & $-1$ & $1$  & $0$  & $0$\\
 \hline\hline
\end{tabular}
\caption{Character table and irreps of the group $\overline{6}m2$ ($D_{3h}$).}
\label{tbl:D3h-group}
\end{table}

\begin{table}[htb]
\begin{tabular}{|c|c|c|}\hline\vspace*{-0.8em}
 & \\
 irrep  &  basis functions  & band\\ 
\hline\vspace*{-0.8em} 
 &  &  \\
\vspace*{-0.8em}
 ${A_1'}$ &  $|{\Psi_{0,0}^{Mo}}\rangle$, $|{\Psi_{2,0}^{Mo}}\rangle$, 
 $\frac{1}{\sqrt{2}}\left(|\Psi_{1,0}^{S1}\rangle-|\Psi_{1,0}^{S2}\rangle\right)$ & VB \\
  &  &  \\  
\hline\vspace*{-0.8em} 
 &  &  \\
\vspace*{-0.8em}
 ${A_2''}$ &  $|{\Psi_{1,0}^{Mo}}\rangle$, $\frac{1}{\sqrt{2}}\left(|\Psi_{1,0}^{S1}\rangle-|\Psi_{1,0}^{S2}\rangle\right)$ & VB-2 \\ 
 &  &\\  
  \hline\vspace*{-0.8em}
 & & \\
 ${E'}$ & $\{ |{\Psi_{2,2}^{Mo}}\rangle, |{\Psi_{2,-2}^{Mo}}\rangle \}$ & VB-3 \\
 \vspace*{-1.0em}
 & & \\
  &  $\left\{ \frac{1}{\sqrt{2}}\left(|\Psi_{1,1}^{S1}\rangle-|\Psi_{1,1}^{S2}\rangle\right), 
  \frac{1}{\sqrt{2}}\left(|\Psi_{1,-1}^{S1}\rangle-|\Psi_{1,-1}^{S2}\rangle\right)\right\}$  &  \\  
  &  &\\
\hline\vspace*{-0.8em}
  & & \\
  $E''$       &   $\{ |{\Psi_{2,1}^{Mo}}\rangle, |{\Psi_{2,-1}^{Mo}}\rangle \}$ & VB-1 \\
  \vspace*{-1.0em}
    & & \\
     &  $\left\{ \frac{1}{\sqrt{2}}\left(|\Psi_{1,1}^{S1}\rangle-|\Psi_{1,1}^{S2}\rangle\right), 
         \frac{1}{\sqrt{2}}\left(|\Psi_{1,-1}^{S1}\rangle-|\Psi_{1,-1}^{S2}\rangle\right) \right\}$
                 &  \\
 & & \\        
\hline\hline
\end{tabular}
\caption{Basis functions for the irreps 
of the small group $D_{3h}$ of the $\Gamma$ point.
$\{\dots\}$ denote the partners of the two-dimensional representations. 
The rightmost column shows that to which band the basis functions contribute.  
}
\label{tbs:bandsymGamma}
\end{table}

Table~\ref{tbl:C3h-group} gives the characters and irreps of $C_{3h}$, pertinent to the  $K$ point and 
Table~\ref{tbs:bandsymK} the  rotating orbitals based Bloch wave function which transform as the irreps of $C_{3h}$.

\begin{table}[htb]
\begin{tabular}{|c|cccccc|}\hline
 $ \overline{6}\, (C_{3h})$ &  $E$  & $C_3$ & $ C_3^{2}$  &  $\sigma_h$  &  $S_3$ & $\sigma_h C_3^2$  \\
 \hline
 $ A'$   & $1$  & $1$        & $1$         & $1$   & $1$         & $1$\\ 
 $A''$   & $1$  & $1$        & $1$         & $-1$  & $-1$        & $-1$\\
 $E_1'$  & $1$  & $\omega$   & $\omega^2$  & $1$   & $\omega$    & $\omega^2$\\
 $E_2'$  & $1$  & $\omega^2$ & $\omega$    & $1$   & $\omega^2$  & $\omega$\\
 $E_1''$ & $1$  & $\omega$   & $\omega^2$  & $-1$  & $-\omega$   & $-\omega^2$\\
 $E_2''$ & $1$  & $\omega^2$ & $\omega$    & $-1$  & $-\omega^2$ & $-\omega$\\
 \hline\hline
\end{tabular}
\caption{Character table of the group $\overline{6}$ ($C_{3h}$).}
\label{tbl:C3h-group}
\end{table}

\begin{table}[thb]
\begin{tabular}{|c|c|c|}\hline\vspace*{-0.8em}
 & \\
\vspace*{-0.8em}
irrep  &  basis functions  & band\\
 & & \\
\hline\vspace*{-0.8em} 
 &  &  \\
\vspace*{-0.8em}
 ${A'}$ &   $|{\Psi_{2,-2}^{Mo}}\rangle$, 
 $\frac{1}{\sqrt{2}}\left(|\Psi_{1,-1}^{S1}\rangle+|\Psi_{1,-1}^{S2}\rangle\right)$ & VB \\
  &  &  \\  
\hline\vspace*{-0.8em} 
 &  &  \\
\vspace*{-0.8em}
 ${A''}$ &  $|{\Psi_{2,1}^{Mo}}\rangle$, $\frac{1}{\sqrt{2}}\left(|\Psi_{1,-1}^{S1}\rangle-|\Psi_{1,-1}^{S2}\rangle\right)$ & CB+1 \\ 
 &  &\\  
  \hline\vspace*{-0.8em}
 & & \\
 ${E_1'}$ & $|{\Psi_{2,0}^{Mo}}\rangle$, $\frac{1}{\sqrt{2}}\left(|\Psi_{1,1}^{S1}\rangle+|\Psi_{1,1}^{S2}\rangle\right)$ & CB \\
 \vspace*{-1.0em}
 & & \\
\hline\vspace*{-0.8em}
  & & \\
  $E_2'$       &   $ |{\Psi_{2,2}^{Mo}}\rangle$, $\frac{1}{\sqrt{2}}\left(|\Psi_{1,0}^{S1}\rangle-|\Psi_{1,0}^{S2}\rangle\right)$ & VB-3 \\
  \vspace*{-1.0em}
    & & \\
     &  & CB+2 \\
  &  &\\  
  \hline\vspace*{-1.0em}
 & & \\   
 $E_1''$ & $|{\Psi_{1,0}^{Mo}}\rangle$,  $\frac{1}{\sqrt{2}}\left(|\Psi_{1,1}^{S1}\rangle-|\Psi_{1,1}^{S2}\rangle\right)$ &  VB-2\\
 & & \\        
 \hline\vspace*{-0.8em}
 & & \\   
 $E_2''$ & $|{\Psi_{2,-1}^{Mo}}\rangle$,  $\frac{1}{\sqrt{2}}\left(|\Psi_{1,0}^{S1}\rangle+|\Psi_{1,0}^{S2}\rangle\right)$ &  VB-1\\
 &  & \\
\hline\hline
\end{tabular}
\caption{Basis functions for the irreducible representations of the small group of the $K$ point.
The rightmost column shows that to which band the basis functions contribute.
The basis functions for the $K'$ point can be obtained by complex-conjugation.
}
\label{tbs:bandsymK}
\end{table}

\section{Bilayer graphene Hamiltonian}
\label{sec:bilayer}

The $\vect{k}\cdot\vect{p}$ Hamiltonian of bilayer graphene\cite{mccann} at the $K$ point of the BZ, 
in the basis of $\{A2 ,B1, A1, B2\}$ sites  is given by
\begin{equation}
 H_{\vect{k}\vect{p}}^{BG}=\left(
 \begin{array}{cccc}
  0 & \gamma_1 & v_4 p_{+} & v_0 p_{-} \\
  \gamma_1  & 0 & v_0 p_{+} & v_4 p_{-} \\
  v_4 p_{-}  & v_0 p_{-} & 0 & v_3 p_{+}\\
  v_0 p_{+} & v_4 p_{+} & v_3 p_{-} & 0 \\
 \end{array}
 \right)
 \label{BGHam}
\end{equation}
where we have chosen the on-site energies to be zero, $p_{\pm}=p_x\pm i p_y$, 
the velocities $v_0$ , $v_3$, and $v_4$ depend on intra- and interlayer hoppings, and 
$\gamma_1$ is the direct hopping between the atoms $B1$ and $A2$. 
One can perform a unitary transformation which 
rotates the Hamiltonian (\ref{BGHam}) into the basis $\{|\Psi_{A1}\rangle,  |\Psi_{A2}\rangle, 
|\Psi_{E_1}\rangle,  |\Psi_{E_2}\rangle \}$
where the basis functions $|\Psi_{\mu}\rangle$ transform as the irreps $\mu=\{A1, A2, E\}$ 
of the small group of the $K$ point, which is $D_3$ in this case. One finds   
\begin{equation}
H_{\vect{k}\vect{p}}^{BG}=
\left(
\begin{array}{cccc}
  -\gamma_1 & 0 & -\tilde{v}_{04}\, p_{+} & \tilde{v}_{04}\, p_{-} \\
  0  & \gamma_1 & \overline{v}_{04}\,  p_{+} & \overline{v}_{04}\, p_{-} \\
    -\tilde{v}_{04}\, p_{-} & \overline{v}_{04}\,  p_{-} & 0 & v_3 p_{+}\\
   \tilde{v}_{04}\, p_{+} & \overline{v}_{04}\, p_{+} & v_3 p_{-} & 0 \\
 \end{array}
 \right)
 \label{BGHamtr}
\end{equation}
where $\tilde{v}_{04}=\frac{1}{\sqrt{2}}(v_0-v_4)$, and $\overline{v}_{04}=\frac{1}{\sqrt{2}}(v_0+v_4)$. 
This Hamiltonian is characterized by the three hoppings $v_3$, $\tilde{v}_{04}$, and $\overline{v}_{04}$, 
and three band edge energies $-\gamma_1$, $\gamma_1$, and $0$ (degenerate). The well known low-energy 
effective Hamiltonian of bilayer graphene\cite{mccann} can be obtained by projecting out the  
states $|\Psi_{A1}\rangle$ and   $|\Psi_{A2}\rangle$. 

The Hamiltonian of monolayer $\textnormal{MoS}_2$ has the same structure as (\ref{BGHamtr}) but is 
characterized by five different hoppings and four different band-edge energies; in this sense it is
a generalization of (\ref{BGHamtr}).

\section{Q-point minimum in the  conduction band}
\label{sec:qpoint}

\begin{figure}[ht]
 \includegraphics[scale=0.52]{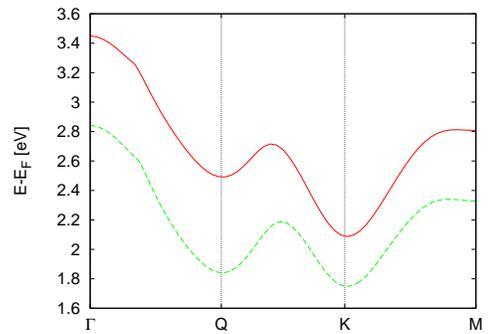}
 \caption{ Conduction band of $\textnormal{MoS}_2$ from DFT calculations using 
 the HSE06 functional (red, solid line) and the LDA (green, dashed).}
 \label{fig:Qpoint}
\end{figure}

In this section we briefly discuss whether it is important to consider the minimum at the 
$Q$ point in the conduction band (see Fig.~\ref{fig:Qpoint}).

To this end we compare the conduction band calculated  in local density approximation (LDA) and using 
the HSE06 hybrid functional\cite{hybridfunct}. The latter, while being computationally much more demanding than the
LDA, has been shown to improve the accuracy of electronic structure calculations for many semiconductors\cite{janesko}.

As one can see in Fig.~\ref{fig:Qpoint}, there are two main differences between the results of calculated 
with the HSE06 functional (red, solid line) and LDA (green, dashed line). 
Firstly,  there is an up-shift of  the HSE06 conduction band with respect to the LDA one, leading to a 
larger band gap at $K$. 
Secondly, the minimum at the $Q$ point is  much higher in energy (and becomes shallower) than the minimum at  $K$ 
in the case of HSE06 calculations.
In particular, the difference between the minima is $E_{Q}^{\rm hybrid}-E_{K}^{\rm hybrid}=0.405\, {\rm eV}$ for 
HSE06 and  $E_{Q}^{\rm LDA}-E_{K}^{\rm LDA}=0.09\, {\rm eV}$ for LDA. 
(Note, that the LDA calculations of Ref.~[\onlinecite{jacobsen}] give $\approx 0.2 \, {\rm eV}$).   
For comparison, the difference between the valence band maxima are 
$E_{K}^{\rm hybrid}-E_{\Gamma}^{\rm hybrid}=0.058\, {\rm eV}$ for 
HSE06  and $E_{K}^{\rm LDA}-E_{\Gamma}^{\rm LDA}=0.12\, {\rm eV}$ for LDA. 
Therefore, regarding transport properties, for p-doped samples states at the  $\Gamma$ point are more 
important than the states at $Q$  for the n-doped case. 

We note that both the increase of the band gap at the $K$ point and the up-shift of the minimum at the Q point 
are in qualitative agreement with the GW calculations of Ref.~[\onlinecite{lambrecht}]. The importance of 
the $Q$ point minimum can hopefully be determined when more accurate measurements of mobility 
 become  available, because the phonon-limited mobility depends quite sensitively on the 
energy separation of $E_{K}-E_{Q}$ (for details see Refs.~[\onlinecite{jacobsen,kim}]).

In contrast, the energy difference between the top of the valence band at the $\Gamma$ and $K$ points shows
much smaller dependence  on the choice of the computational method.

\section{$\mathbf{k\cdot p}$ parameters from calculations with HSE06 hybrid functional}
\label{hse-bandparam}

Comparison between experimental data and DFT calculations suggest\cite{blaha} that in the case of semiconductors 
hybrid functionals\cite{hybridfunct} not only produce band gaps which are in better agreement with 
 measurements but also the calculated effective masses are closer to the experimental values. 
Motivated by this we have also fitted our model to LDA band structure calculations
performed with the HSE06 functional\cite{viktor2}. The main effect at the $K$ point  seems to be that 
the effective masses become lighter and the coupling parameter $\gamma_3$ stronger. 
However, the change in the effective mass at the $\Gamma$ point is more significant. 
In Table \ref{bandparams} we show the relevant band parameters calculated both with  
LDA and using HSE06. 
\begin{table}[hb]
 \begin{tabular}{|c|c|c|}\hline
    -- &  LDA  & HSE06 \\
 \hline
 $\alpha$ &  $ 1.73\, {\rm eV \AA}^2 $ &  $ 1.57 \, {\rm eV \AA}^2 $  \\
 \hline
 $\beta$  &   $ -0.13\, {\rm eV \AA}^2 $&   $ 0.1 \, {\rm eV \AA}^2 $ \\
 \hline
 $\gamma_3$  & $3.82\, {\rm eV \AA}$ &      $ 4.13 \, {\rm eV \AA}$   \\
 \hline
 $\kappa$  &  $ -1.02\, {\rm eV \AA}^2 $ &  $ -1.12\, {\rm eV \AA}^2 $ \\
 \hline
 $\eta$  &  $8.53\, {\rm eV \AA}^3 $ &  $  7.87\, {\rm eV \AA}^3 $ \\
 \hline
 $m_{\rm eff}^c/m_e $  &  $0.48$ & $0.43$ \\
 \hline
 $m_{\rm eff}^v/m_e$  &  $-0.62$  &   $-0.53$ \\
 \hline
 $m_{\rm eff}^{\Gamma}/m_e$ & $-3.65$ & $-2.24$ \\ 
 \hline
 \end{tabular}
 \caption{Parameters of the effective Hamiltonian. $m_e$ is the bare electron mass.}
 \label{bandparams}
\end{table}

\newpage

\bibliographystyle{prsty}

\begin{thebibliography}{10}
\bibitem{viktor}
The plane-wave cutoff energy was 600 eV. We used a $12 \times 12$ 
Monkhorst-Pack k-point grid in the 2D plane to relax the structure and a 
$24 \times 24$ grid to calculate the band structure.
The optimised lattice parameter was $3.129 \AA{}$  and the distance between 
the two layers of S atoms was $3.114 \AA{}$.
\bibitem{neil-derivative}
The matrix elements of the momentum operator were evaluated  within the 
LDA  using the CASTEP  code\cite{clark}, because 
 the necessary plane-wave coefficients of the wave functions at the band edges
 were readily accessible in the output of CASTEP. 
We used norm-conserving pseudopotentials, a plane-wave cutoff energy of 80 Ha
 an artificial periodicity of 30 Bohr in the vertical direction and a $21 \times 21$ 
Monkhorst-Pack mesh. The optimised lattice parameter and the S-S distance was the
same as in the case of VASP calculations, see Ref.\cite{viktor}.
\bibitem{viktor2}
The plane-wave cutoff energy was 600 eV and 
we used 24x24 Monkhorst-Pack k-point  grid to 
calculate the band structure.
\end{thebibliography}

\end{document}